\documentclass[fleqn,12pt]{article}
\usepackage[numbers,sort&compress]{natbib}
\usepackage[footnotesep=0.4in]{geometry}
\usepackage{graphicx}
\usepackage{caption2}
\usepackage{amsmath}
\usepackage{bbding}
\usepackage{graphics}
\usepackage{pifont}
\usepackage{amsfonts}
\usepackage{caption2}
\usepackage{float}
\makeatletter
\newcommand\figcaption{\def\@captype{figure}\caption}
\newcommand\tabcaption{\def\@captype{table}\caption}
\DeclareMathOperator{\sech}{Sech}
 \topmargin
-0.8in \textheight 9.4in \textwidth6.8in \hoffset -1.45cm

\begin{document}

\date{}
\title{Darboux Transformation for the Nonisospectral and Variable-coefficient KdV Equation}
\author{Ling-Jun Liu,   Xin Yu\thanks{Corresponding
author, with e-mail address as yuxin@buaa.edu.cn}
\\{\em Ministry-of-Education Key Laboratory of Fluid Mechanics and
National}\\
{\em  Laboratory for Computational Fluid Dynamics,  Beijing University of }\\
{\em  Aeronautics and Astronautics, Beijing 100191, China} }

\maketitle

\vspace{8mm}

\begin{abstract}
With the nonuniform media taken into account, the nonisospectral and
variable-coefficient Korteweg-de Vries equation, which describes
various physical situations such as fluid dynamics and plasma, is
under investigation in this paper. With appropriate selection of
wave functions, the Darboux transformation is constructed, by which
the multi-soliton solutions are derived and graphs are presented.
The spectral parameters, coefficients and initial phase are
discussed analytically and numerically to demonstrate their
respective effect on the soliton dynamics, which plays a role in
achieving the feasible soliton management with explicit conditions
taken into account.

\end{abstract}
\vspace{3mm}


\vspace{3mm}


\vspace{20mm}

\newpage
\noindent {\Large{\bf I. Introduction}}

 \vspace{3mm}

The Korteweg-de Vries-type (KdV-type) equations are of current
interest in describing the physical situations such as wave motion,
ion-acoustic wave, Bose-Einstein condensates and arterial
dynamics~\cite{hd1,hd2,bt,yanzy,xuejk,gg,lisc,duanws}. By
appropriate model construction, KdV-type equations with certain
coefficients have been investigated analytically by inverse
scattering method~\cite{wlchan3,wlchan2}, direct
method~\cite{xuxiaoge2,lilili}, Wronskian
technique~\cite{dengshufang,xuxiaoge1}, Darboux
transformation~\cite{hhc,xll,wxy,sj,lqp} and so
on~\cite{zhaoxiqiang,wlchan1,dingshuangshuang}, among which, the
Darboux transformation (DT) is a purely algebraic iterative tool to
generate soliton solutions~\cite{hhc,xll,wxy,sj,lqp}.

It is shown that an integrable system may possess more than one
DT~\cite{hhc} and so far various forms of DT have been investigated
for different kinds of KdV-type equations. For instance, a more
general DT is constructed for the Hirota-Satsuma-KdV equation, which
is quadratic in the spectral parameter~\cite{xll}, the DT with an
arbitrary parameter is utilized for the study of integrable
sixth-order KdV equation~\cite{wxy}, a vectorial DT with ordinary
determinants is applied for the supersymmetric extension of KdV
equation~\cite{lqp}, and the generalized binary DT is used to obtain
the negaton, positon, and complexiton solutions of nonisospectral
KdV equations with self-consistent sources~\cite{sj}.

With the nonuniform media taken into account, the KdV-type equations
are related to time-dependent spectral parameters with
relaxation~\cite{hirota3,zhangdajun,hhh,ntk,lq,gmr,zhangjiefang}:

\begin{equation}\label{equation}
\hspace{10mm}u_t\,+a(t)\,u\,u_{x}+b(t)\,u_{xxx}+c(t)\,u+[d_1(t)+d_2(t)x]\,u_{x}=0\,,
\end{equation}

under appropriate coefficient selections which may model for the
problem of tunnelling of solitons across density humps~\cite{gmr}
and indicate the reflectionless potentials for nonisospectral KdV
hierarchy~\cite{ntk}. With a certain balancing between the loss and
the nonuniformity in media~\cite{hirota3}, the nonisospectral
characteristics of the motion behaviours of some solutions are
described~\cite{zhangdajun} and the symmetry structure is
studied~\cite{zhangjiefang}. Taking self-consistent sources in
nonlinear media into account, Eq.~\ref{equation} can be extended and
investigated~\cite{lq,hhh}.

By virtue of the constraint below~\cite{yuxincc}, which reduce the
degrees of restriction among various coefficients,
Eq.~(\ref{equation}) is an integrable system,

\begin{equation}\label{cc}
\hspace{10mm}a(t)=\frac{6\,b(t)}{\rho}\,e^{\int{\![c(t)-2d_2(t)]dt}}\,,
\end{equation}
where $\rho$ is a non-zero constant.

However, to the best of our knowledge, the DT for
Eq.~(\ref{equation}) has not been constructed under the general
constraint~(\ref{cc}). Hereby in this paper, we will construct a DT
of Eq.~(\ref{equation}) and provide analytical and numerical
discussion about the solitonic dynamic under the effect of spectral
parameters, coefficients and initial phases respectively.

In Section II, the DT for Eq.~(\ref{equation}) under
constraint~(\ref{cc}) is constructed. In Section III, the soliton
solutions are derived in explicit forms, with the soliton-like
solutions being generated and presented graphically. In Section IV,
we investigate the effects of spectral parameters, coefficients and
initial phases on the solitonic dynamic, as well as present the
graphic illustration of the soliton propagation and interactions. In
Section V, numerical simulation is applied to further discuss the
soliton activities with the constraint~(\ref{cc}) perturbed.
Finally, the conclusions are offered in Section VI.

\vspace{7mm} \noindent {\Large{\bf II. Construction of DT}}

\vspace{3mm} In this section, we will construct a DT of
Eq.~(\ref{equation}) under the Lax pairs~\cite{yuxincc}
\begin{align}\label{lax1}
&\hspace{-3mm}\Psi_t+\Big[d_1(t)+d_2(t)x+6\chi
\,b(t)e^{-2\int{\!d_2(t)dt}}+3\lambda (t)
\,b(t)\Big]\Psi_x+b(t)\Psi_{xxx}\notag\\
\
&\hspace{0mm}+3\,b(t)\Big[\frac{1}{\rho}\,e^{\int{\![c(t)-2d_2(t)]dt}}u
-\chi e^{-2\int{\!d_2(t)dt}}\Big]\Psi_x=0\,,
\end{align}
\begin{equation}\label{lax2}
\hspace{-1.5mm}\Psi_{xx}+\Big[\frac{1}{\rho}\,e^{\int{\![c(t)-2d_2(t)]dt}}u
-\chi e^{-2\int{\!d_2(t)dt}}\Big]\Psi-\lambda(t)\Psi=0\,,
\end{equation}
where $\Psi$ is defined as wave function with $\rho\neq0$ and $\chi$
being a real constant. The spectral parameter $\lambda(t)$ satisfies
the conditions of Lax pair by $2\,d_2(t)\lambda(t)+\lambda'(t)=0$.

The DT can be constructed with undetermined time-varying
coefficients A(t), B(t) and C(t) as below,
\begin{equation}\label{dt1}
\hspace{10mm}u_N=u_{N-1}+A(t)\,\frac{\partial^2}{\partial
x^2}\,\log{\Phi_{N-1}}\,,
\end{equation}
and
\begin{equation}\label{dt2}
\hspace{10mm}\Psi_N=B(t)\,\frac{\partial}{\partial
x}\,\Psi_{N-1}-C(t)\,\Psi_{N-1}\,\frac{\partial}{\partial
x}\,\log{\Phi_{N-1}}\,,
\end{equation}
in which subscript $N$ represents $N-th$ iteration of DT and the
$\Psi_{N}$ satisfying Lax pairs is a real function of $x$ and $t$.
The particular solution $\Phi_{N}(x,t)$ is defined as
$\Phi_{N}(x,t)=\Psi_{N}(x,t,\lambda_{N}(t))$.

Define the $\Phi_{N}(x,t)$ as an induction function for its role in
the transformation process. Notice that $\Psi_0$ and $u_0$ are the
original solutions of the Lax pairs and Eq.~(\ref{equation}),
respectively, with $\Psi_1$ and $u_1$ derived from as new solutions
by a DT.

Spectral parameter $\lambda(t)$ is a time-varying function, which
should comply with the conditions of the Lax pair and is given by
\begin{equation}\label{condition1}
\hspace{10mm}\lambda(t)=e^{\int{\![-2d_2(t)]dt}}\,\overline{\lambda}\,,
\end{equation}
where $\overline{\lambda}$ is a real constant, reflecting the size
of value in the spectral parameter $\lambda(t)$ when $d_2(t)$ is
given. In the discussion later, we study the spectral parameter
$\lambda(t)$ by discussing the value of $\overline{\lambda}$.

Applying the undetermined expressions~(\ref{dt1}) and~(\ref{dt2})
into the following compatibility condition of Lax pair~(\ref{lax1})
and~(\ref{lax2})

\begin{equation}\label{e1}
\hspace{10mm}\Psi_{Ntxx}(x,t)=\Psi_{Nxxt}(x,t).,
\end{equation}
and finish it by comparative coefficients method, we obtain the
specific coefficients of DT~(\ref{dt1}) and~(\ref{dt2})
\begin{equation}\label{A1}
\hspace{10mm}A(t)=2\,\overline{\lambda}\,\rho\,e^{\int{\![2d_2(t)-c(t)]dt}}\,,
\end{equation}
\begin{equation}\label{B1}
\hspace{10mm}B(t)=\overline{\lambda}\,e^{\int{\![d_2(t)]dt}}\,,
\end{equation}
\begin{equation}\label{C1}
\hspace{10mm}C(t)=\overline{\lambda}\,e^{\int{\![d_2(t)]dt}}\,.
\end{equation}

Therefore, we can construct the explicit form of DT as following,
\begin{equation}\label{dtf1}
\hspace{10mm}u_N=u_{N-1}+2\,\overline{\lambda}\,\rho\,e^{\int{\![2d_2(t)-c(t)]dt}}\,\frac{\partial^2}{\partial
x^2}\,\log{\Phi_{N-1}}\,,
\end{equation}
and
\begin{equation}\label{dtf2}
\hspace{10mm}\Psi_N=e^{\int{\!d_2(t)dt}}\,\frac{\partial}{\partial
x}\,\Psi_{N-1}-e^{\int{\!d_2(t)dt}}\,\Psi_{N-1}\,\frac{\partial}{\partial
x}\,\log{\Phi_{N-1}}\,,
\end{equation}
in which $\Phi_{N-1}(x,t)$=$\Psi_{N-1}(x,t,\lambda_{N-1}(t))$ and
$u_N$ is derived from $u_{N-1}$ by an iteration of DT. Notice that
each step of DT utilizes a specified parameter $\lambda_{N}(t)$ for
derivation of new solution.

\vspace{7mm} \noindent {\Large{\bf II. Generations by DT}}

\vspace{3mm}In this section, we will discuss the generated solutions
for Eq.~(\ref{equation}) under different forms of wave functions by
DT.

Employ $u_0=0$ as the original solution for Eq.~(\ref{equation}) and
suppose that the original solution of wave function has the
following form
\begin{equation}\label{f1}
\hspace{10mm}\Psi_0(x,t)=e^{k(t)\,x+w(t)}+e^{-[k(t)\,x+w(t)]}\,,
\end{equation}
where $k(t)$ and $w(t)$ are functions of $t$ to be determined.

Substituting above equation into Lax pair~(\ref{lax1})
and~(\ref{lax2}), we have
\begin{equation}\label{k1}
\hspace{10mm}k(t)=\sqrt{\overline{\lambda}+\chi}\,\,e^{\int{\!-d_2(t)dt}}\,,
\end{equation}
and
\begin{equation}\label{w1}
\hspace{10mm}w(t)=c-\int{[\!\,4\,({\overline{\lambda}+\chi})^{\frac{3}{2}}\,b(t)\,e^{\int{\!-3\,d_2(t)}dt}+e^{\int{\!-\,d_2(t)dt}}]dt}\,,
\end{equation}
in which $c$ is an arbitrary constant, representing the soliton
initial phase.

By means of expression~(\ref{f1}), an induction solution $\Phi_0$ is
constructed by $\Phi_{0}(x,t)$=$\Psi_{0}(x,t,\lambda_{0}(t))$ with a
given value $\overline{\lambda_0}$. Hereby, we obtain a
single-soliton solution by a DT~(\ref{dtf1}).

We have
\begin{equation}\label{u1}
\hspace{10mm}u_1(x,t)=2\,\sqrt{\chi+\overline{\lambda_0}}\,e^{\int{-c(t)dt}}\,\sech^{2}{T(x,t)}\,,
\end{equation}
where
\begin{equation}\label{T}
\hspace{5mm}T(x,t)=c_1+\rho\,\sqrt{\chi+\overline{\lambda_0}}\,[x\,e^{\int{-\,d_2(t)dt}}-\int{\!U(t)dt}]\\,
\end{equation}

\begin{equation}\label{U}
\hspace{10mm}U(t)=e^{\int{-3\,d_2(t)dt}}[4\,(\chi+\overline{\lambda_0})\,b(t)+e^{\int{2\,d_2(t)dt}}]\\,
\end{equation}
in which $c_1$ is the initial phase with the subscript representing
the first iteration of DT.

To derive wave function $\Psi_1$, we employ a different wave
function defined as
\begin{equation}\label{psi0}
\hspace{10mm}\Psi_{0}^{'}(x,t)=e^{k^{'}(t)\,x+w^{'}(t)}-e^{-[k^{'}(t)\,x+w^{'}(t)]}\,,
\end{equation}
where $k^{'}(t)$ and $w^{'}(t)$ are as same as the
expression~(\ref{k1}) and~(\ref{w1}) respectively. Therefore, we can
obtain another wave function by a DT~(\ref{dtf2}) using
$\Psi_{0}^{'}$ and $\Phi_0$, which paves the way for deriving
two-soliton solution.

\begin{equation}\label{psi1}
\hspace{10mm}\Psi_1=e^{\int{\![d_2(t)]dt}}\,\frac{\partial^2}{\partial
x}\,\Psi_{0}^{'}-e^{\int{\![d_2(t)]dt}}\,\Psi_{0}^{'}\,\frac{\partial^2}{\partial
x}\,\log{\Phi_0}\,.
\end{equation}

To obtain two-soliton solution, we utilize a second DT~(\ref{dtf1})
with an induction solution $\Phi_1$ corresponding to the wave
function~(\ref{psi1}),
\begin{equation}\label{u2}
\hspace{10mm}u_2=u_1+2\,\rho\,\overline{\lambda}\,e^{\int{\![2d_2(t)-c(t)]dt}}\,\frac{\partial^2}{\partial
x^2}\,\log{\Phi_1}\,,
\end{equation}
where $\Phi_1(x,t)=\Psi_1(x,t,\lambda_1(t))$. The derived
two-soliton comply with the basic law of nature in its propagation
and interaction dynamic. An example is presented below in Fig.1 (a).

By the induction solution corresponding to Eq.~(\ref{f1}) without
adjustment, the solutions for Eq.~(\ref{equation}) can also be
generated, but with considerable singularity in rational field. As
shown in Fig.1 (b), the soliton-like graphs are presented.

\begin{minipage}{\textwidth}
\renewcommand{\captionfont}{\scriptsize}
\renewcommand{\captionlabelfont}{\scriptsize}
\renewcommand{\captionlabeldelim}{.\,}
\renewcommand{\figurename}{Fig.\,}
\hspace{2.3cm}\includegraphics[scale=0.55]{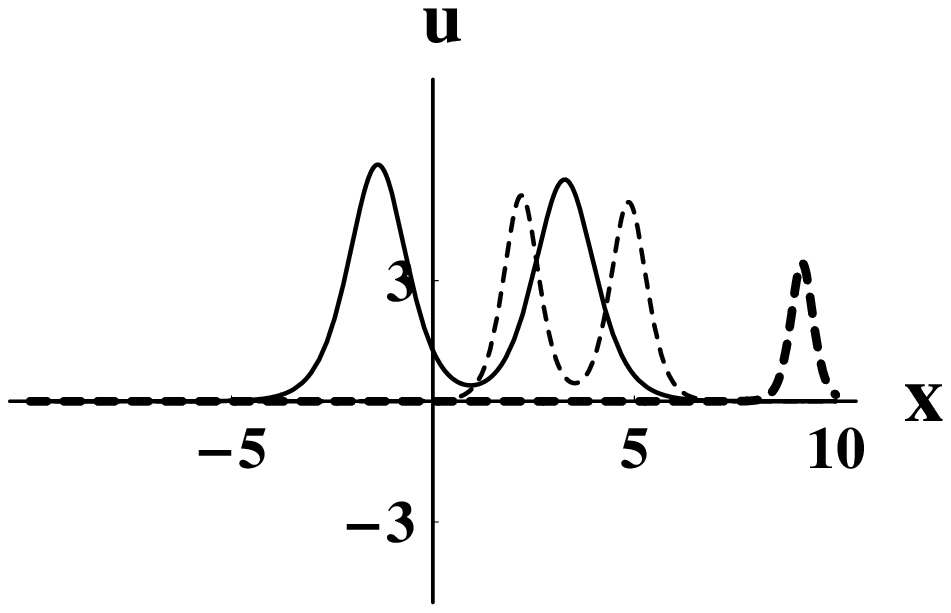}
\hspace{1.5cm}\includegraphics[scale=0.55]{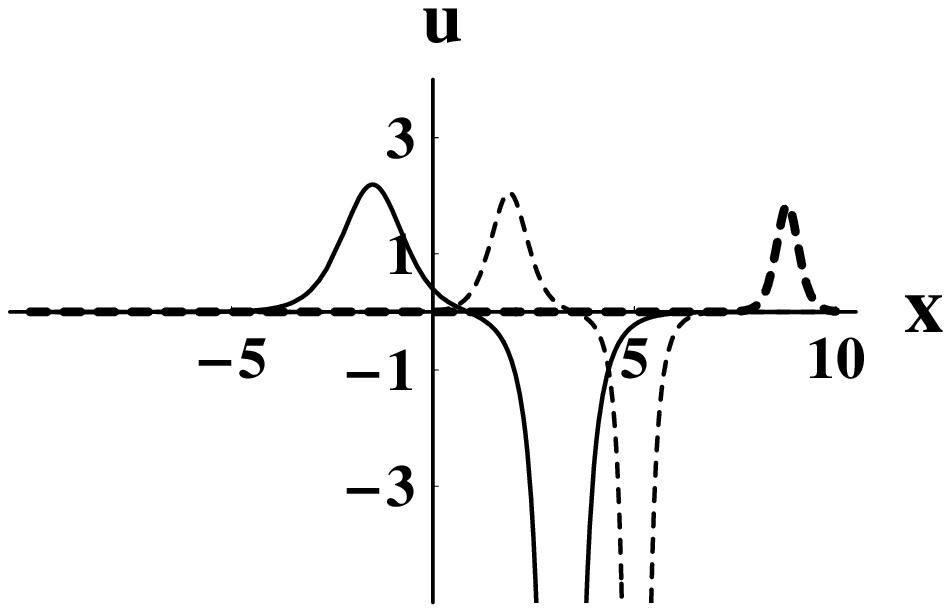}
\vspace{-0.5cm}{\center\hspace{4.7cm}\footnotesize ($a$)
\hspace{6.6cm}($b$)} \figcaption{ The solutions by DT with adjusted
wave function(a) and non-adjusted wave function(b) where
coefficients and parameters are $\rho=1$, $\chi=0$,
$\overline{\lambda_0}=1$, $\overline{\lambda_1}=1.1$, $b(t)=1$,
$c(t)=Sin[t]$, $d_1(t)=0$, $d_2(t)=-1$ and $c_1=1, c_2=0$. when
$t=0$ (solid line), $t=0.5$ (dashed line), $t=1.0$ (bold dashed
line).} \label{tu1}
\end{minipage}
\\[\intextsep]

The difference between the generated solutions derived by the
original wave function~(\ref{f1}) and the adjusted wave
function~(\ref{psi0}) indicates the necessary modification when
constructing DT in the form of operator in order to derive soliton
solutions. As shown in Fig.2 (a) and (b), the solution iterated by
DT can be further extended to multi-soliton solution.

\begin{minipage}{\textwidth}
\renewcommand{\captionfont}{\scriptsize}
\renewcommand{\captionlabelfont}{\scriptsize}
\renewcommand{\captionlabeldelim}{.\,}
\renewcommand{\figurename}{Fig.\,}
\hspace{2.3cm}\includegraphics[scale=0.55]{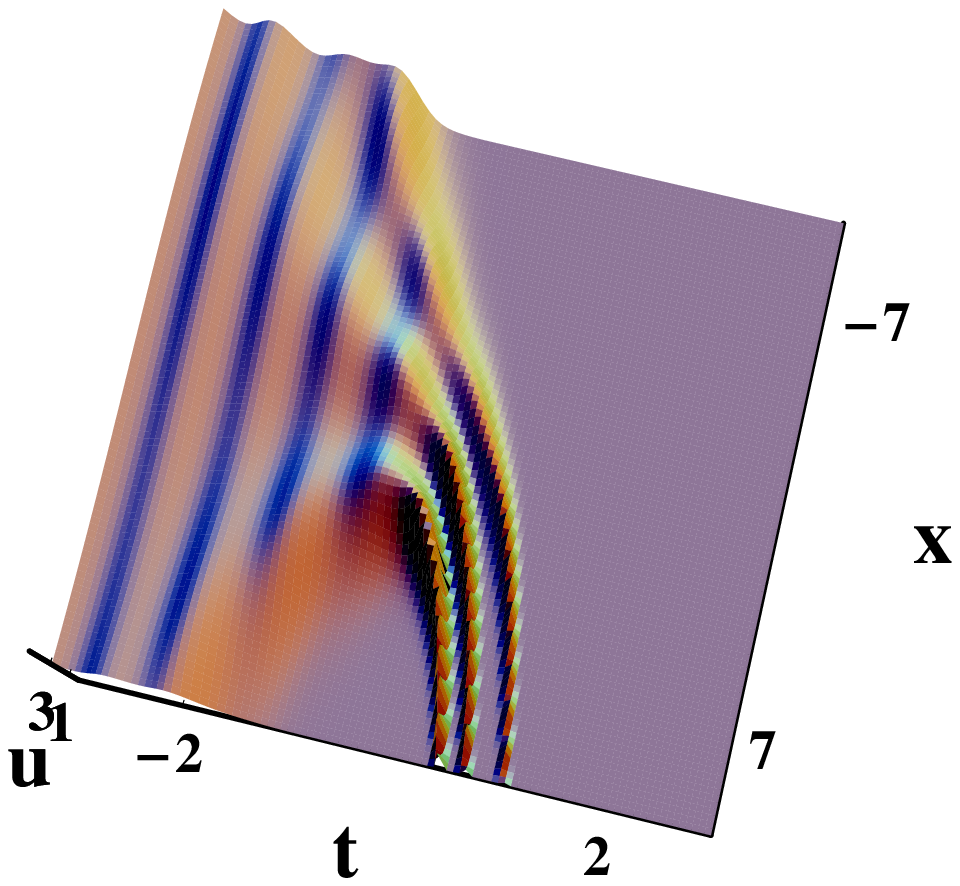}
\hspace{1.5cm}\includegraphics[scale=0.55]{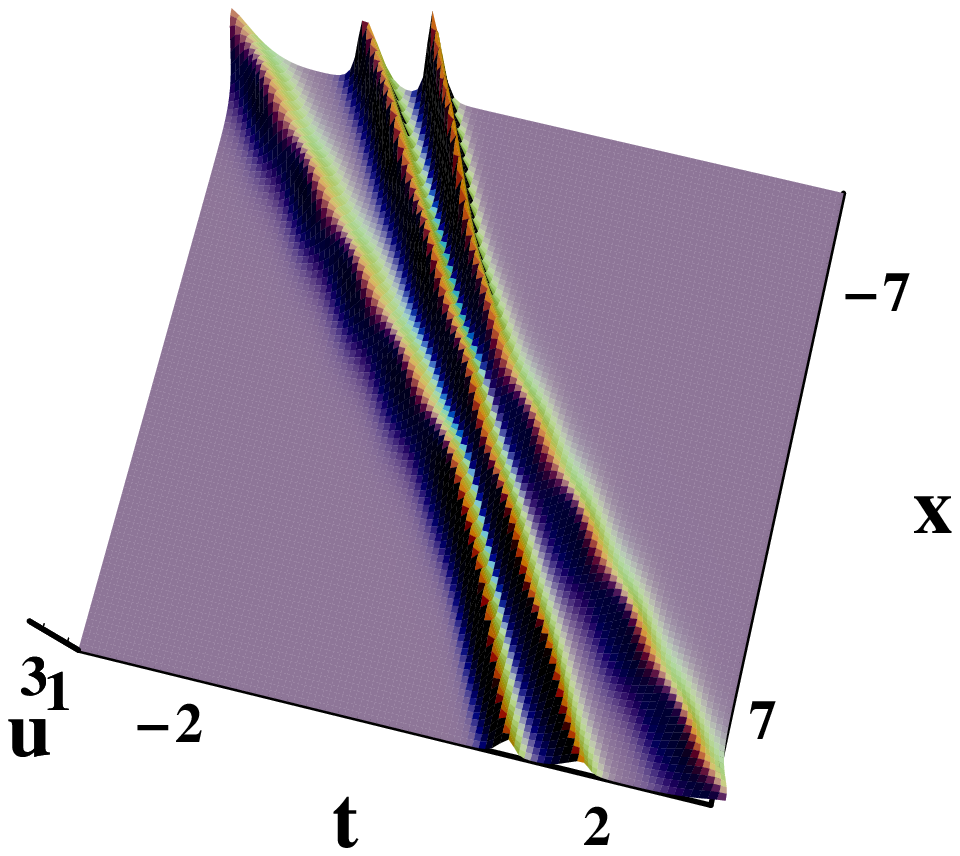}
\vspace{-0.5cm}{\center\hspace{4.7cm}\footnotesize ($a$)
\hspace{6.6cm}($b$)} \figcaption{Three-soliton solution by DT
(a)$\rho=1$, $\chi=0$, $\overline{\lambda_0}=1$,
$\overline{\lambda_1}=2$, $\overline{\lambda_2}=3$, $b(t)=1$,
$c(t)=Sin[10t]$, $d_1(t)=0.1\,t^3$, $d_2(t)=-1$ and
$c_1=c_2=c_3=0$.(b)$\rho=1$, $\chi=0$, $\overline{\lambda_0}=1$,
$\overline{\lambda_1}=1.5$, $\overline{\lambda_2}=2$, $b(t)=1$,
$c(t)=Sin[10t]$, $d_1(t)=0$, $d_2(t)=0$ and $c_1=c_2=c_3=0$.}
\label{tu1}
\end{minipage}
\\[\intextsep]

\vspace{7mm} \noindent {\Large{\bf IV. Discussions}}

\vspace{3mm} Generally speaking, the interaction and propagation of
soliton dynamics play a role in determining the physical feature and
benefit the applicants of dynamical system. Therefore, in this
section, we will investigate the respective influence of spectral
parameters, different coefficients and initial phases on the
solitonic velocity, width, amplitude, demonstrating the figure of
solitonic dynamic in fluid.

\vspace{3mm}\emph{Case A.} Eq.~(\ref{equation}) is related to the
time-dependent spectral parameter $\lambda(t)$, which could describe
the solitons wave in nonhomogeneous media~\cite{hd1,hd2}.

\begin{minipage}{\textwidth}
\renewcommand{\captionfont}{\scriptsize}
\renewcommand{\captionlabelfont}{\scriptsize}
\renewcommand{\captionlabeldelim}{.\,}
\renewcommand{\figurename}{Fig.\,}
\hspace{0.5cm}\includegraphics[scale=0.65]{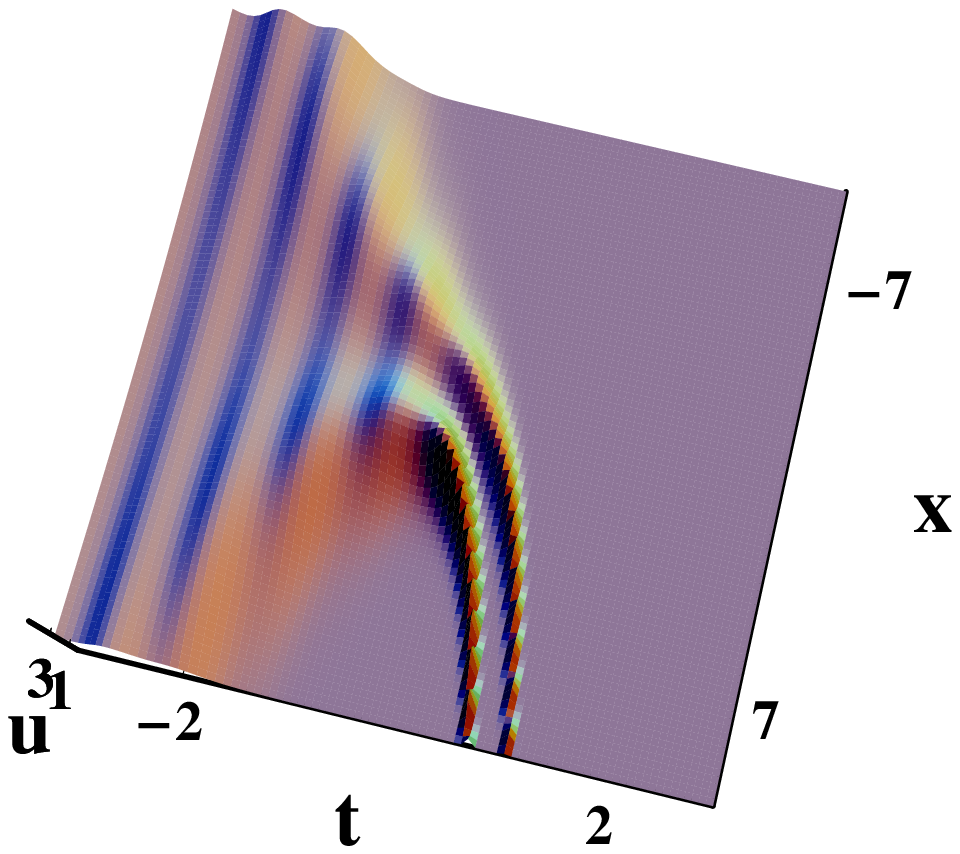}
\hspace{1.5cm}\includegraphics[scale=0.65]{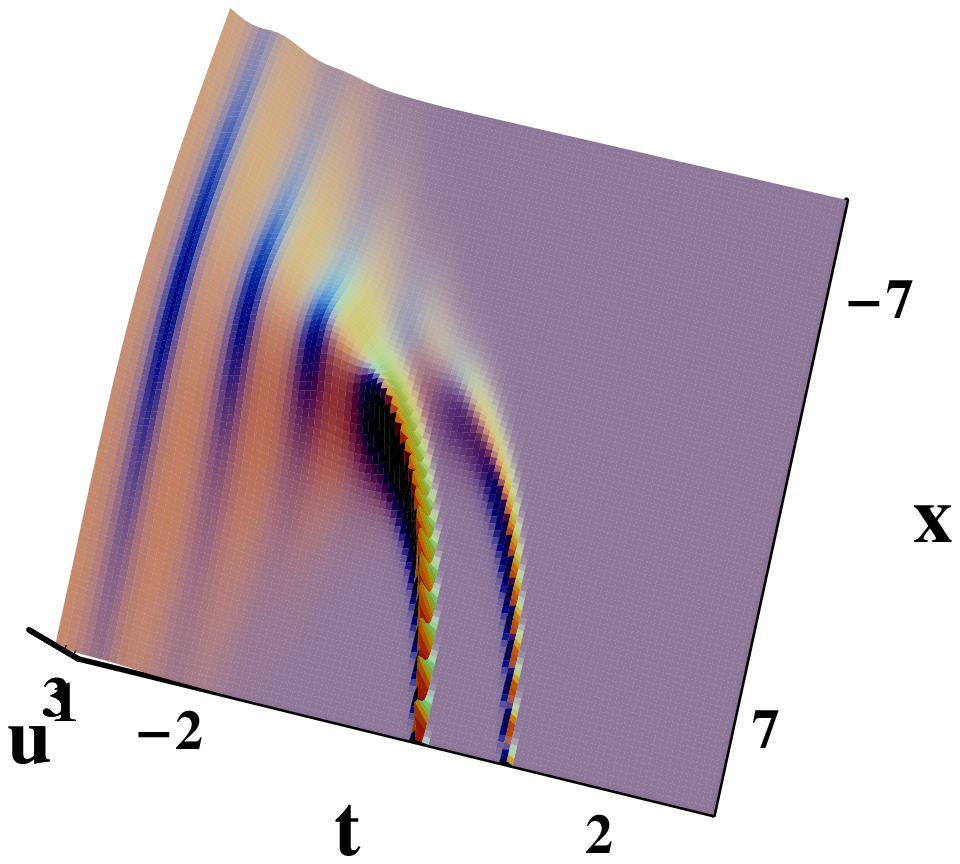}
\vspace{-0.5cm}{\center\hspace{3.3cm}\footnotesize ($a$)
\hspace{7.7cm}($b$)}

\hspace{0.5cm}\includegraphics[scale=0.65]{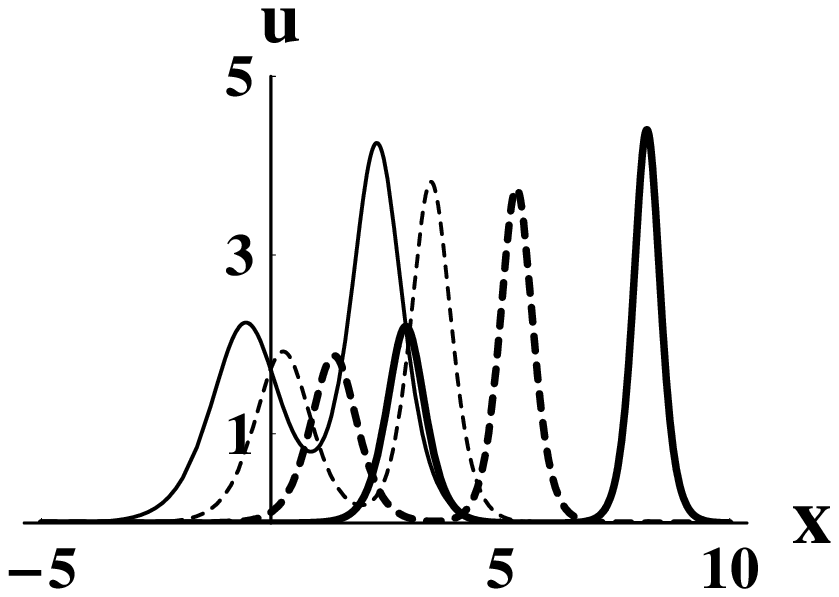}
\hspace{1.5cm}\includegraphics[scale=0.65]{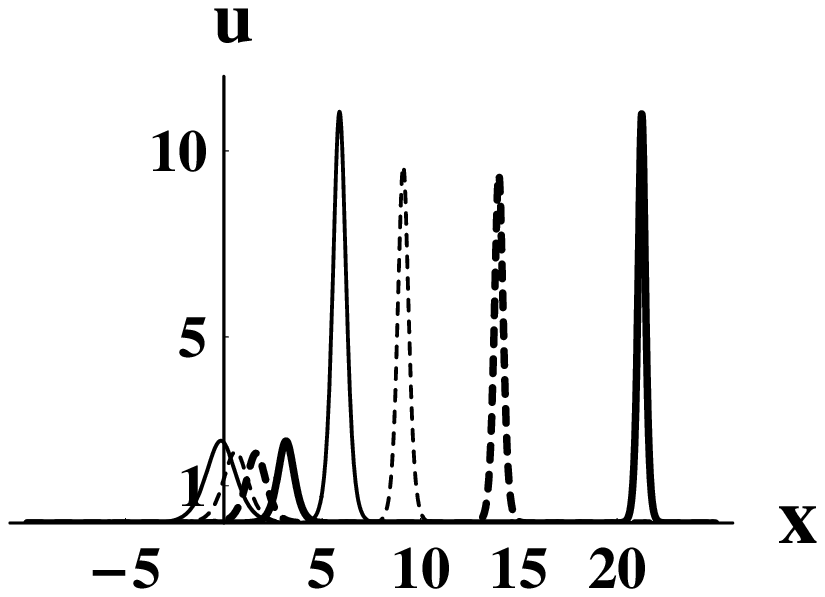}
\vspace{-0.5cm}{\center\hspace{3.3cm}\footnotesize ($c$)
\hspace{7.7cm}($d$)} \figcaption{Solitonic propagation and
interaction with coefficients $\rho=1$, $\chi=0$, $b(t)=1$,
$c(t)=sin(10t)$, $d_1(t)=d_2(t)=-1$ and $c_1=0,c_2=0$. The spectral
parameters: (a)$\overline{\lambda_0}=1$, $\overline{\lambda_1}=2$
(b) $\overline{\lambda_0}=1$, $\overline{\lambda_1}=5$. (c) and (d)
Profiles (a) and (b) respectively at different time: (c) t=0 (solid
line), t=0.2 (dashed line), t=0.4 (bold dashed line), t=0.6 (bold
solid line); (d) t=0 (solid line), t=0.2 (dashed line), t=0.4 (bold
dashed line), t=0.6 (bold solid line).}\label{tu3}
\end{minipage}
\\[\intextsep]

As shown in Figs.3 (a) and (b), when spectral parameter
$\overline{\lambda_1}$ increases, the interaction of solitons become
intensifier, with the amplitude aggraded and the wide narrowed.
Consider Figs.3 (a), (c) and (b), (d), respectively, with time
passing by, the solitons propagate longer distance at the same time
interval. In each graph, the soliton with higher amplitude generates
a faster speed compared with the other one. Compare Figs.3 (a), (c)
with (b), (d), as spectral parameter $\overline{\lambda_1}$
increases, the elevation soliton experiences a growth of velocity
and amplitude. It is shown in Fig.3 (b) that when there is an
noticeable imbalance between spectral parameters, a soliton dynamic
feature is almost covered by the other at the intersection.

It is worth noting that when
$\overline{\lambda_0}=\overline{\lambda_1}$, the two-soliton
solution is degenerated to constant value zero.

By an appropriate selection of the $\chi$, we can construct negative
spectral coefficients $\overline{\lambda}$ in the following graph.
The phenomena in the Fig.4 can be also interpreted by the influence
of spectral parameters. When $\overline{\lambda}$ is balanced with
$\chi$, the solitons converge into one at the intersection.

\begin{minipage}{\textwidth}
\renewcommand{\captionfont}{\scriptsize}
\renewcommand{\captionlabelfont}{\scriptsize}
\renewcommand{\captionlabeldelim}{.\,}
\renewcommand{\figurename}{Fig.\,}
\hspace{1.5cm}\includegraphics[scale=0.65]{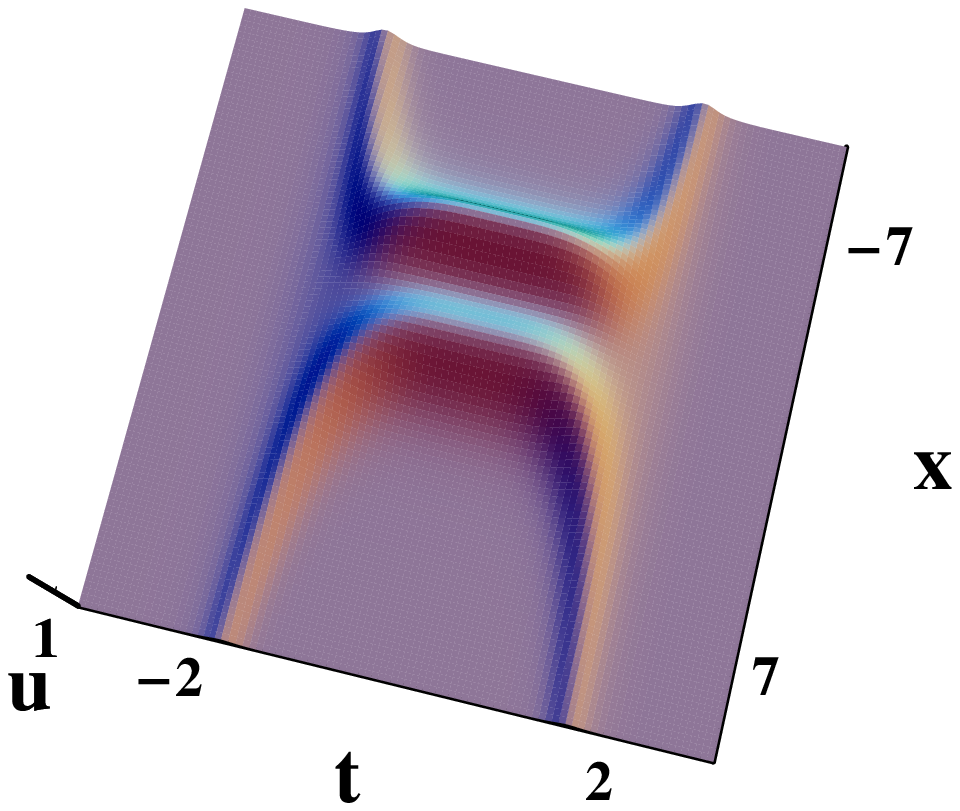}
\hspace{1.5cm}\includegraphics[scale=0.65]{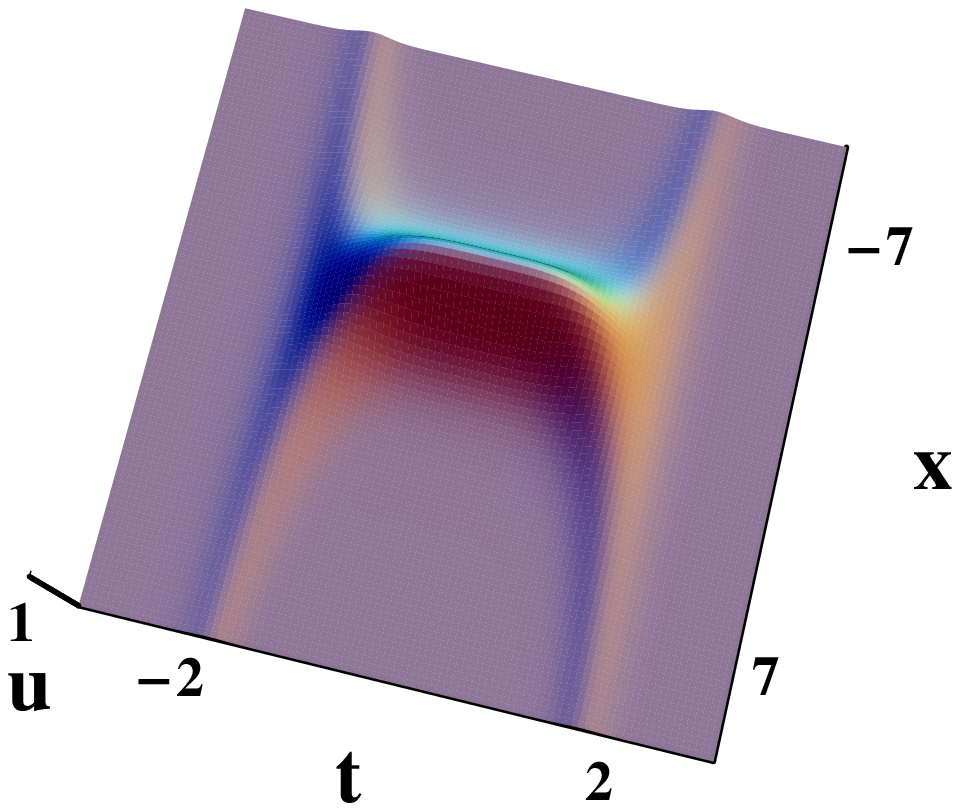}
\vspace{-0.5cm}{\center\hspace{4.7cm}\footnotesize ($a$)
\hspace{7.6cm}($b$)} \figcaption{ Solitonic propagation and
interaction with coefficients $\rho=1$, $\chi=1$,
$b(t)=c(t)=d_1(t)=d_2(t)=t^3$ and $c1=3$, $c2=0$. The spectral
parameter: (a) $\overline{\lambda_0}=-0.5$,
$\overline{\lambda_1}=-0.1$ (b)$\overline{\lambda_0}=-1$,
$\overline{\lambda_1}=-0.5$.}\label{tu8}
\end{minipage}
\\[\intextsep]

\vspace{3mm}\emph{Case B.}

From the expression~(\ref{u1}), the solitonic amplitude can be
interpreted as
$2\,\sqrt{\chi+\overline{\lambda_0}}\,e^{\int{-c(t)dt}}\,$, where
the $c(t)$ is the only time-dependent coefficient.

\begin{minipage}{\textwidth}
\renewcommand{\captionfont}{\scriptsize}
\renewcommand{\captionlabelfont}{\scriptsize}
\renewcommand{\captionlabeldelim}{.\,}
\renewcommand{\figurename}{Fig.\,}
\hspace{1.5cm}\includegraphics[scale=0.65]{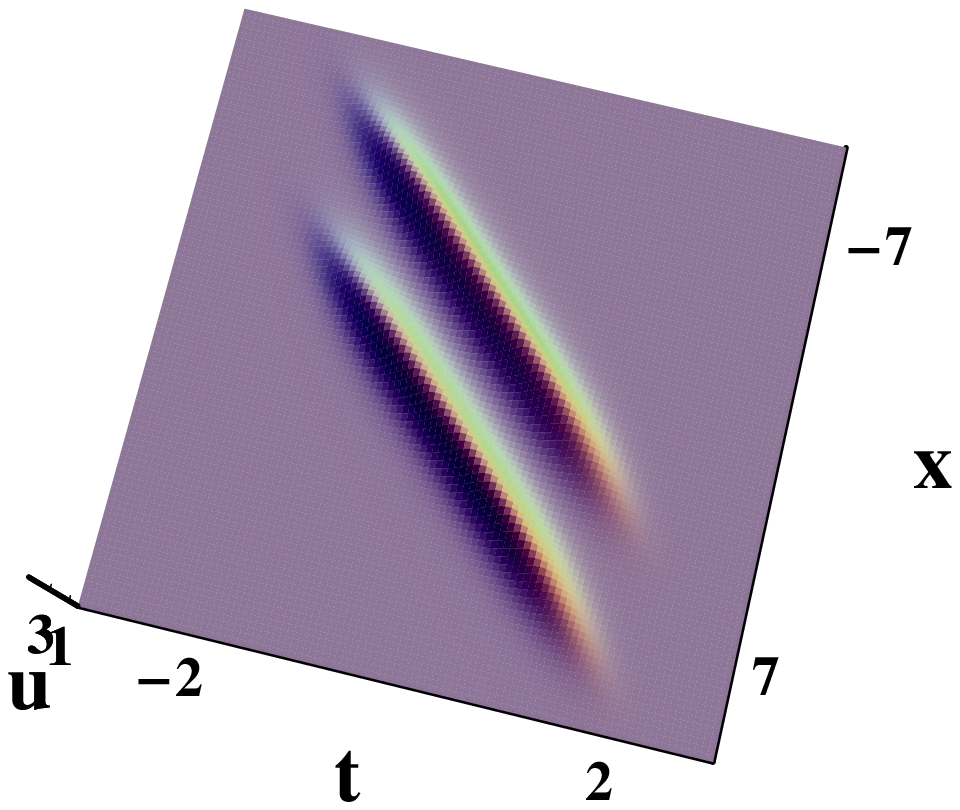}
\hspace{1.5cm}\includegraphics[scale=0.65]{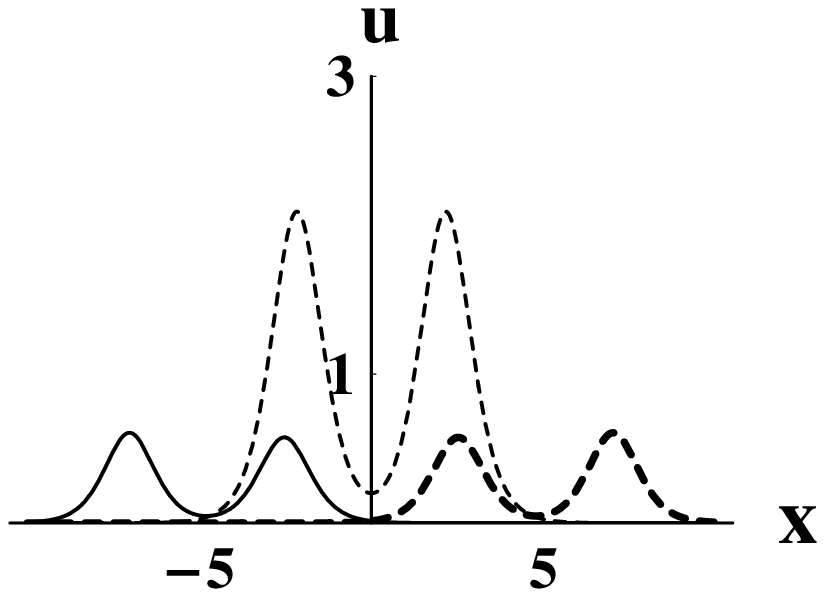}
\vspace{-0.5cm}{\center\hspace{4.7cm}\footnotesize ($a$)
\hspace{7.6cm}($b$)} \figcaption{ Solitonic propagation and
interaction with coefficients and parameters $\rho=1$, $\chi=0$,
$\lambda_0=1$, $\lambda_1=1.1$, $b(t)=1$, $d_1(t)=-1$ ,
$d_2(t)=c_1=c_2=0$, and $c(t)=t^3$. (b) profiles (a) when $t=-1.5$
(dashed line), $t=0$ (solid line), $t=1.5$ (bold dashed line).}
\label{tu2}
\end{minipage}
\\[\intextsep]

By a positive application of line-damping coefficient $c(t)$, the
soliton management for time-space locality is achieved. As
demonstrated in Fig.5 (b) by the profile at certain time, the
solitonic amplitude get enhanced when time is approaching to zero
and then attenuated as time keeps going on, besides, the space
influenced by solitonic dynamics is limited locally. The depression
of solitons amplitude can be interpreted by the term
$2\,\sqrt{\chi+\overline{\lambda_0}}\,e^{\int{-c(t)dt}}$
analytically.

\vspace{3mm}\emph{Case C.}

The coefficient $d_2(t)$ indicates the inhomogeneities of media,
which affects the solitonic dynamical behavior.

\begin{minipage}{\textwidth}
\renewcommand{\captionfont}{\scriptsize}
\renewcommand{\captionlabelfont}{\scriptsize}
\renewcommand{\captionlabeldelim}{.\,}
\renewcommand{\figurename}{Fig.\,}
\hspace{1.5cm}\includegraphics[scale=0.65]{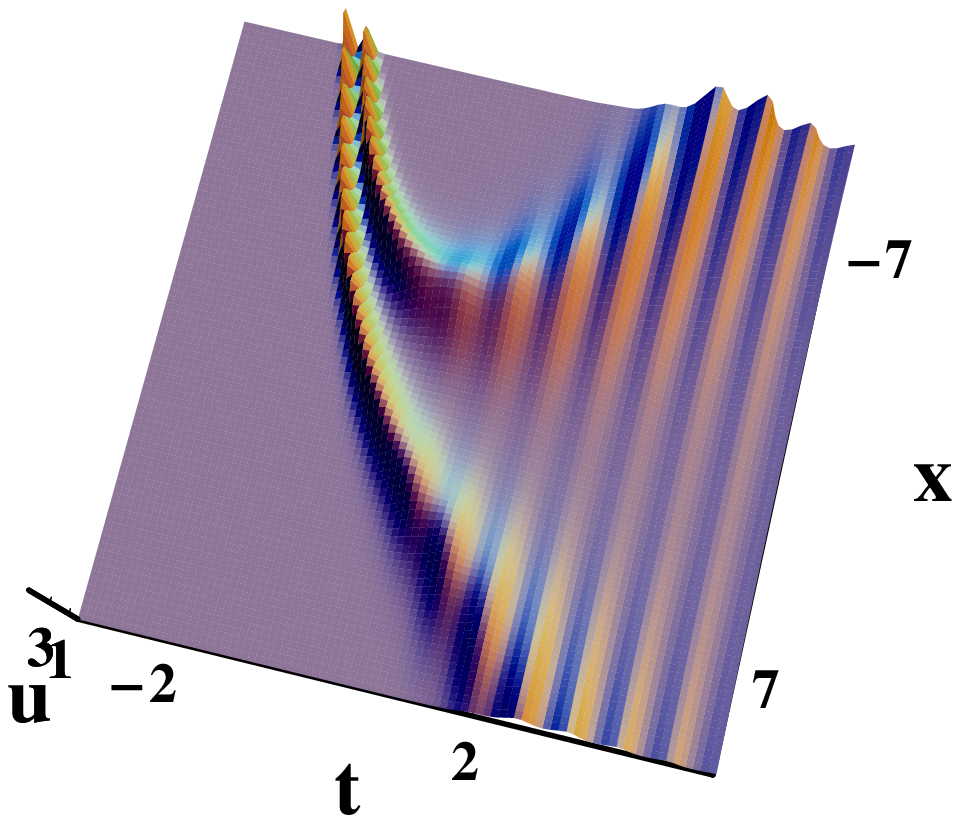}
\hspace{1.5cm}\includegraphics[scale=0.65]{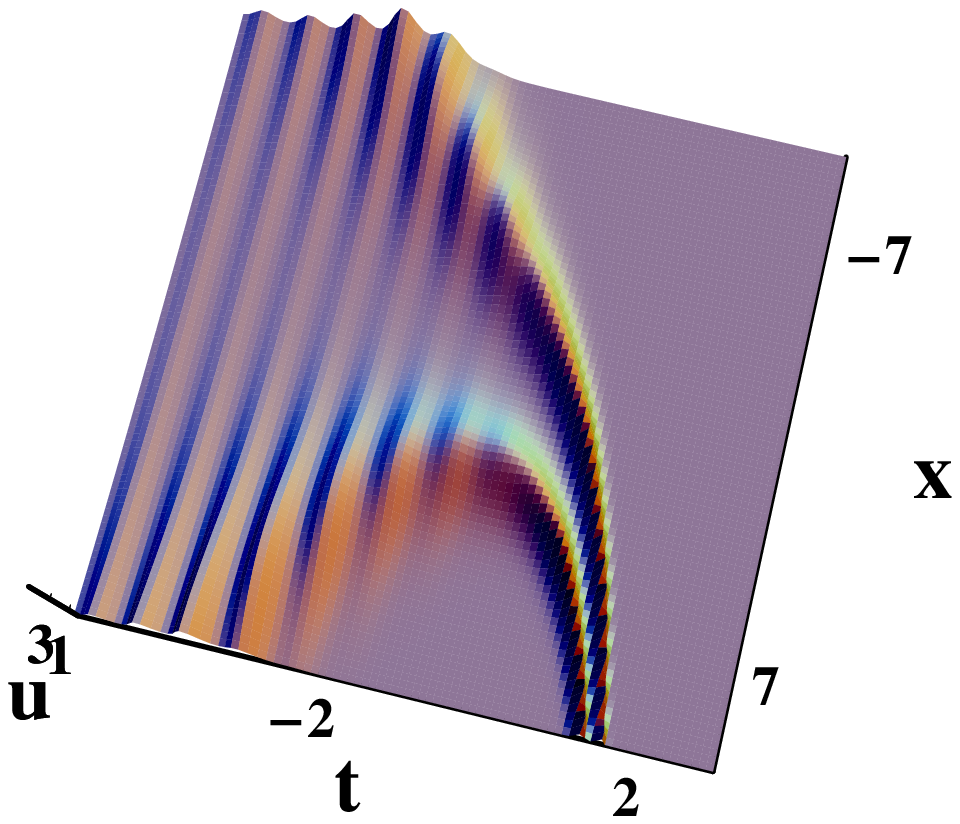}
\vspace{-0.5cm}{\center\hspace{4.7cm}\footnotesize ($a$)
\hspace{7.6cm}($b$)} \figcaption{ Solitonic propagation and
interaction with coefficients and parameters $\rho=1$, $\chi=0$,
$\overline{\lambda_0}=1$, $\overline{\lambda_1}=1.1$, $b(t)=-1$,
$c(t)=t\,sin(10t)$, $d_1(t)=c1=c2=0$. The non-uniformity
coefficient: (a) $d_2(t)=0.5$ (b) $d_2(t)=-0.5$.} \label{tu3}
\end{minipage}
\\[\intextsep]

As shown in Figs.6 (a) and (b), by choosing the negative and
positive value corresponding to the nonuniformity coefficient
$d_2(t)$, the solitons swell and shrink over time, respectively. It
is worth noting that the degree of solitonic broadening and
compressing depends on the absolute value of $d_2(t)$. The influence
of variable coefficients on the soliton dynamics is similar to that
in Refs~\cite{yuxin,yuxincc}.

\vspace{3mm}\emph{Case D.} In case D, based on
expression~(\ref{u2}), the effect of the initial phases will be
discussed and presented graphically.

\begin{minipage}{\textwidth}
\renewcommand{\captionfont}{\scriptsize}
\renewcommand{\captionlabelfont}{\scriptsize}
\renewcommand{\captionlabeldelim}{.\,}
\renewcommand{\figurename}{Fig.\,}
\hspace{0.8cm}\includegraphics[scale=0.4]{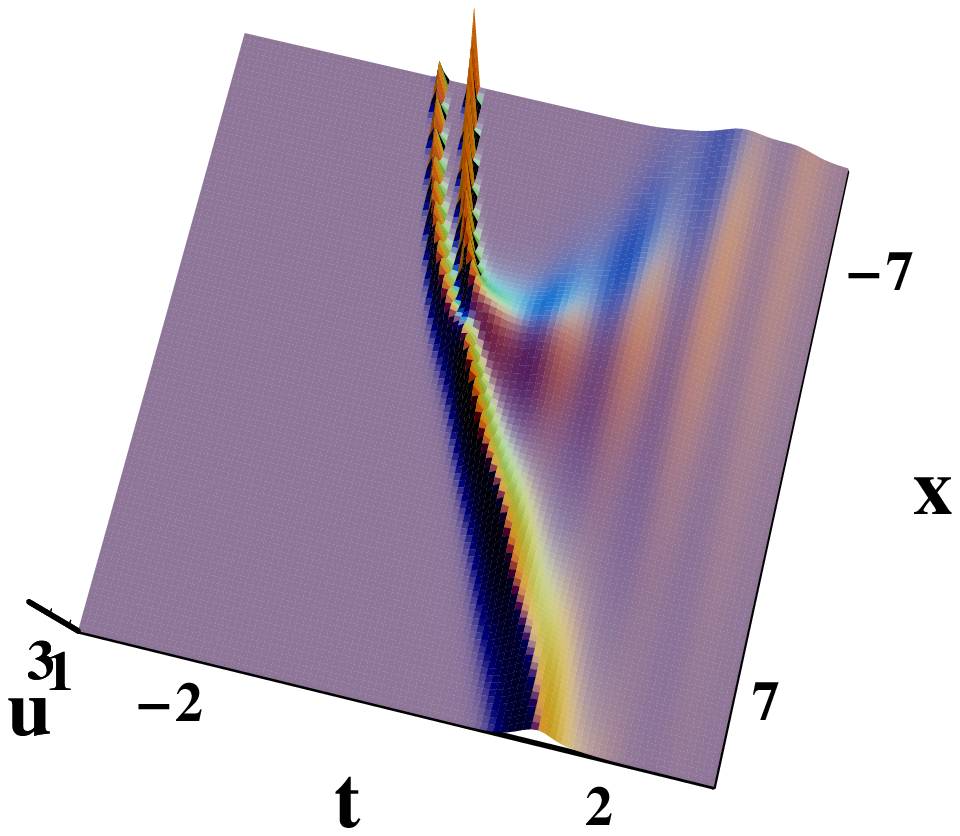}
\hspace{1.2cm}\includegraphics[scale=0.4]{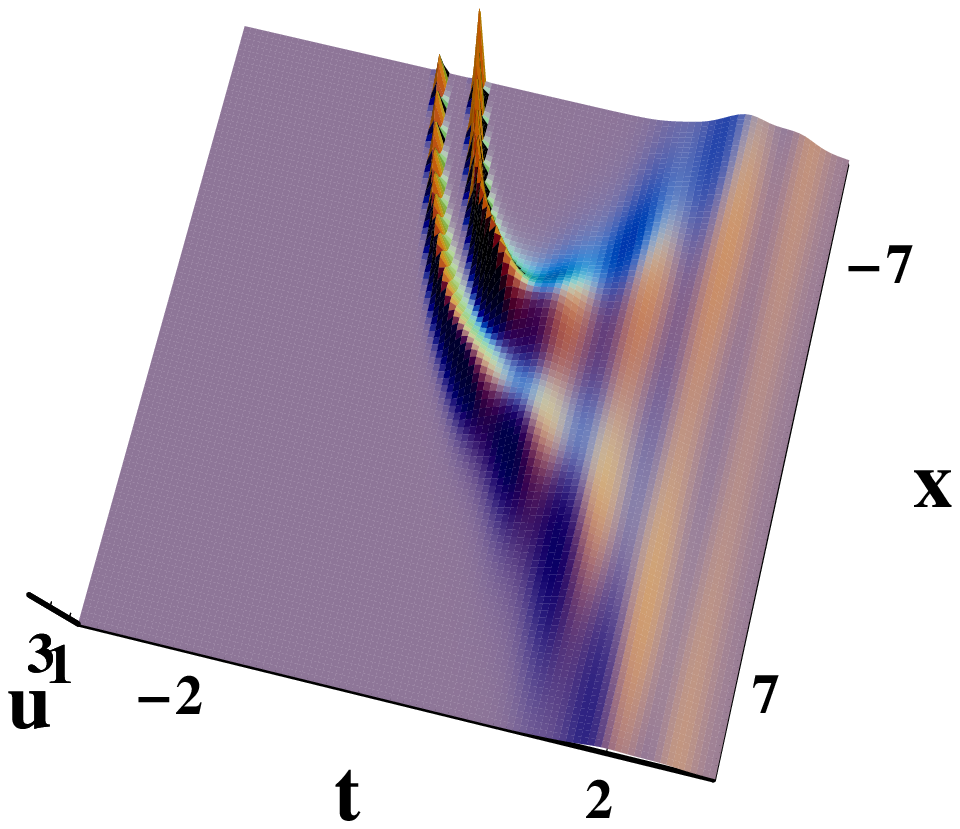}
\hspace{1.2cm}\includegraphics[scale=0.4]{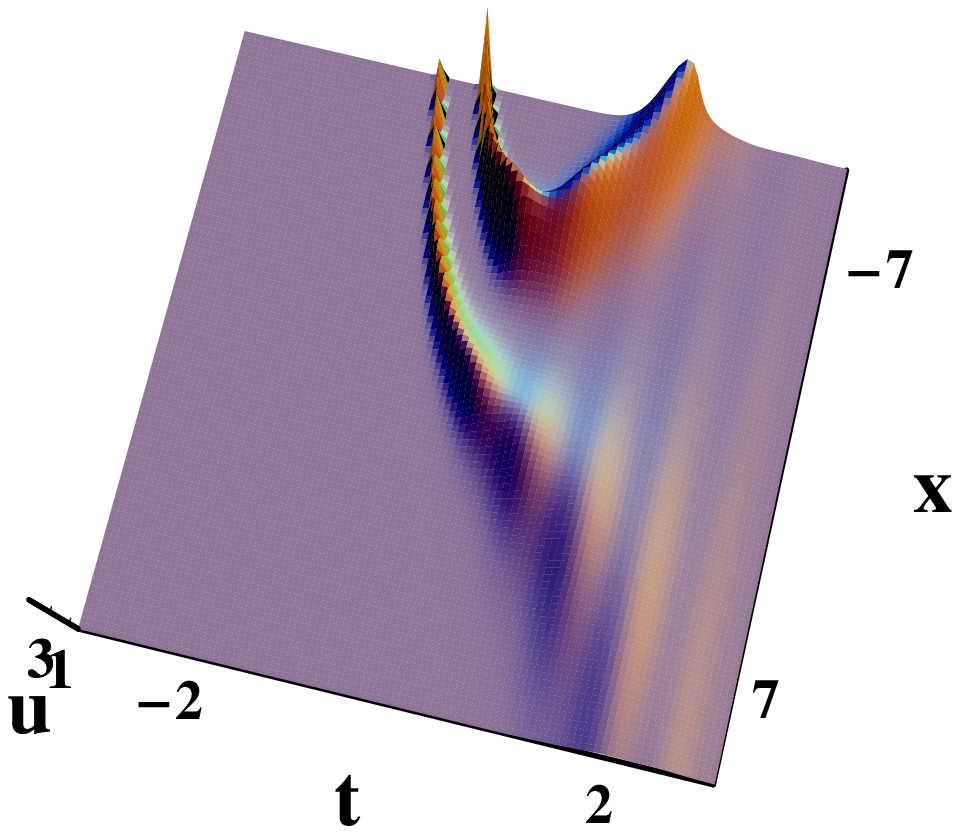}
\vspace{-0.5cm}{\center\hspace{2.2cm}\footnotesize ($a$)
\hspace{5cm}($b$)\hspace{5cm}($c$)} \figcaption{ Solitonic
propagation and interaction with coefficients and parameters
$\rho=1$, $\chi=0$, $\overline{\lambda_0}=1$,
$\overline{\lambda_1}=2$, $b(t)=1$, $c(t)=sin(10t)$, and $d_1(t)=-1,
d_2(t)=1$. The initial phase: (a) $c_1=0,c_2=-3$ (b) $c_1=0,c_2=0$.
(c) $c_1=0,c_2=3$ } \label{tu4}
\end{minipage}
\\[\intextsep]

The initial phase $c_1$ and $c_2$ result from the first and second
step of DT, respectively. As shown in Fig.7 (a), the solitons
interact with each other and generate a phase shift at the
intersection. By an increment of initial phase $c_2$, Figs.7 (a),
(b) and (c) demonstrate a position displacement of the related
soliton, with the other soliton dynamics almost unchanged. By a
proper use of initial phase, we may obtain individual solitonic
management and induce an interaction. With periodical time-dependent
$c(t)=Sin(10\,t)$ in Fig.7, the solitonic physical feature also
corresponds to the interpretation of line-damping coefficient $c(t)$
above, which demonstrates that the periodical amplitude fluctuation
of solitons is generated by the time-periodical value of coefficient
$c(t)$.

\begin{minipage}{\textwidth}
\renewcommand{\captionfont}{\scriptsize}
\renewcommand{\captionlabelfont}{\scriptsize}
\renewcommand{\captionlabeldelim}{.\,}
\renewcommand{\figurename}{Fig.\,}
\hspace{0.5cm}\includegraphics[scale=0.65]{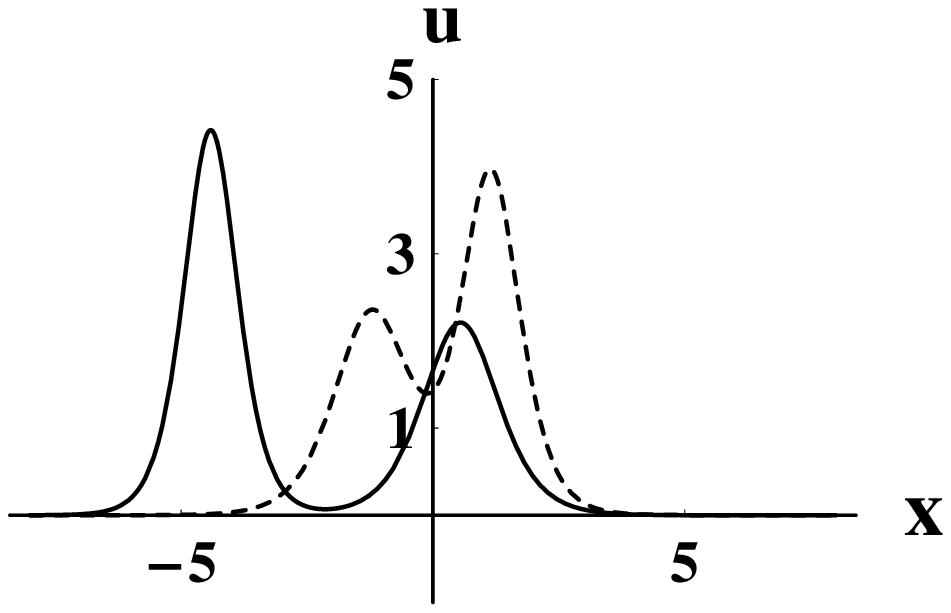}
\hspace{1.5cm}\includegraphics[scale=0.65]{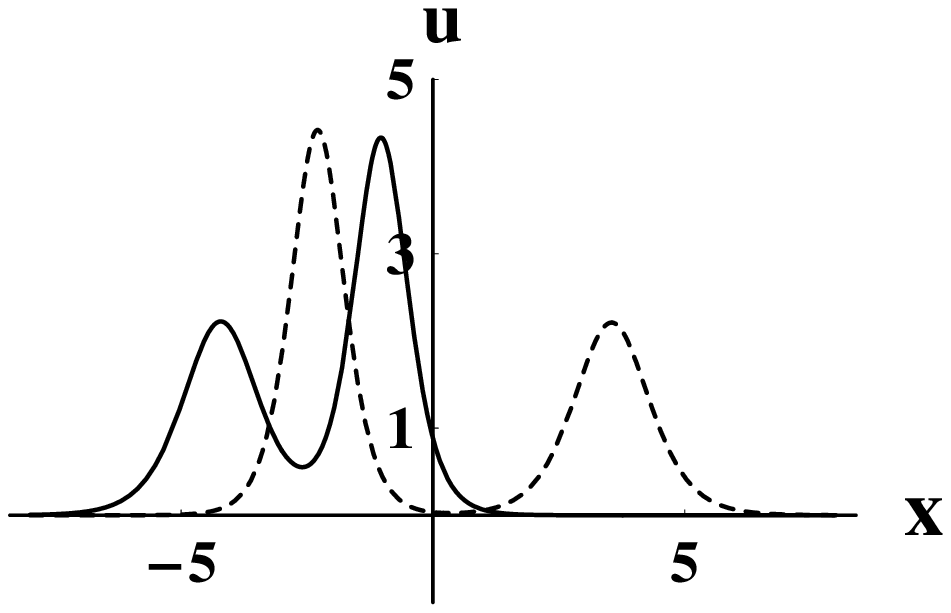}
\vspace{-0.5cm}{\center\hspace{3.3cm}\footnotesize ($a$)
\hspace{7.7cm}($b$)} \figcaption{profiles of Fig.2 at t=0 with
initial phase (a)$c_1=0,c_2=-3$(dashed line) , $c_1=0,c_2=3$ (solid
line); (b)$c_1=-3,c_2=0$(dashed line) , $c_1=3,c_2=0$ (solid
line).}\label{tu3}
\end{minipage}
\\[\intextsep]

It is worth noting that a sign change of the initial phase distance
will lead to an exchange of solitonic position, with the height
nearly unchanged, which is shown in Fig.8. When the initial phase is
appropriately modified, we can generate a solitonic position
exchange in space.

\vspace{7mm} \noindent {\Large{\bf V. Numerical Simulation}}

\vspace{3mm}For solitonic generation, the constraint is necessary to
balance the nonlinearity and dispersion, under which our analytical
solitons-solution is constructed. In this section, when the
constraint~(\ref{cc}) is perturbed, the numerical simulation is
applied to further discuss the influences of coefficients on the
solitonic velocity, amplitude and width.

Employ one-soliton solution~(\ref{u1}) with coefficients and
parameters $b(t)=1, c(t)=2, d_1(t)=1, d_2(t)=1,
\overline{\lambda_0}=0, \chi=1, \rho=1$, we have

\begin{equation}\label{u1_numerical}
\hspace{10mm}u_1(x,t)=2\,e^{-2\,t}\,\sech^{2}{[\frac{4}{3}\,e^{-3\,t}+e^{-t}+x\,e^{-\,t}]}\,,
\end{equation}

Take $u_1(x,0)$ as the initial analytical solution and by finite
difference method, we have the numerical solution
$\widetilde{u_1(x,1)}$.

By $\pm 30\% $ perturbation on the coefficients $b(t)$ $c(t)$,
$d_1(t)$ and $d_2(t)$, respectively, we have the numerical
simulations presented below.

\begin{minipage}{\textwidth}
\renewcommand{\captionfont}{\scriptsize}
\renewcommand{\captionlabelfont}{\scriptsize}
\renewcommand{\captionlabeldelim}{.\,}
\renewcommand{\figurename}{Fig.\,}
\hspace{0cm}\includegraphics[scale=0.5]{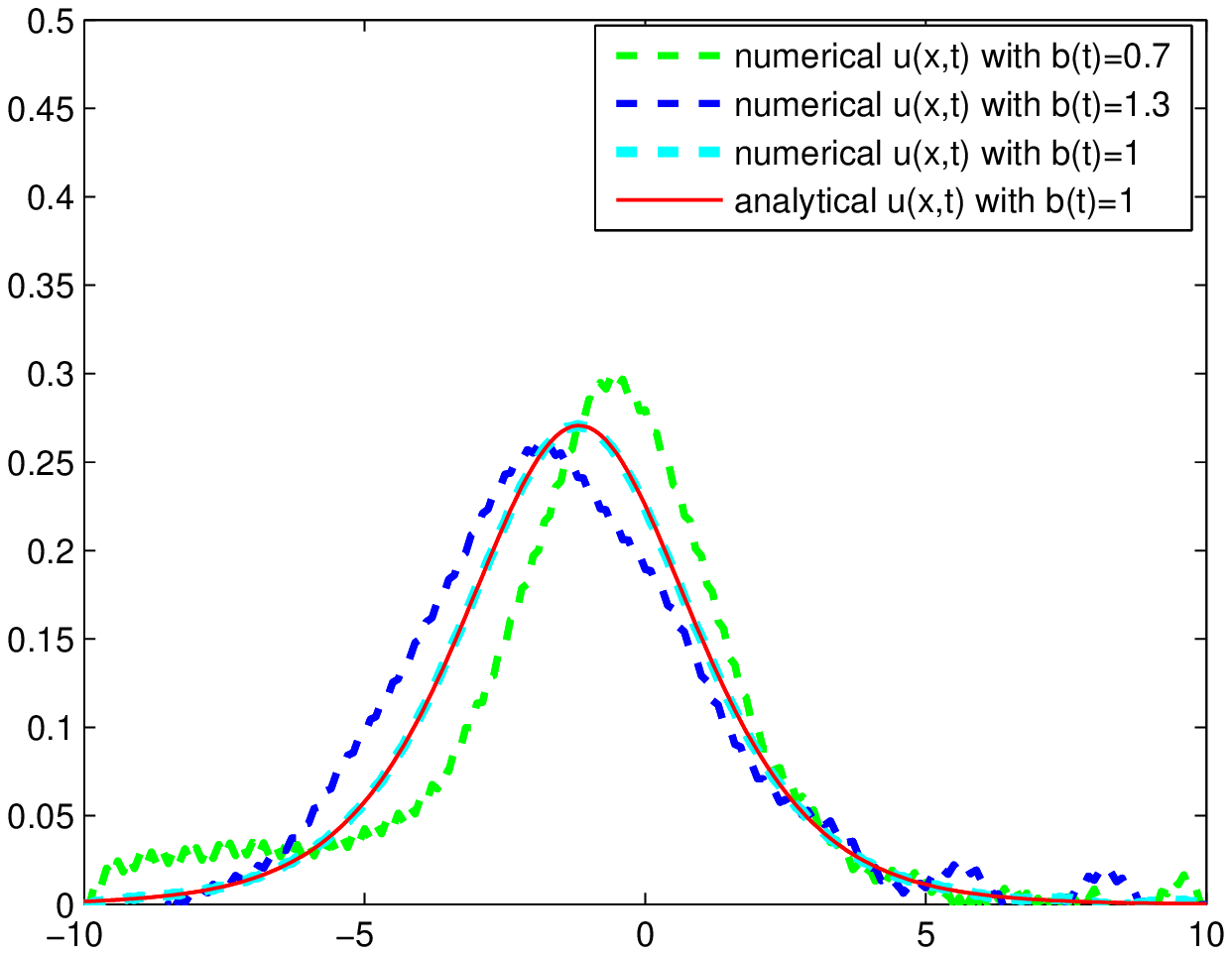}
\hspace{1.5cm}\includegraphics[scale=0.5]{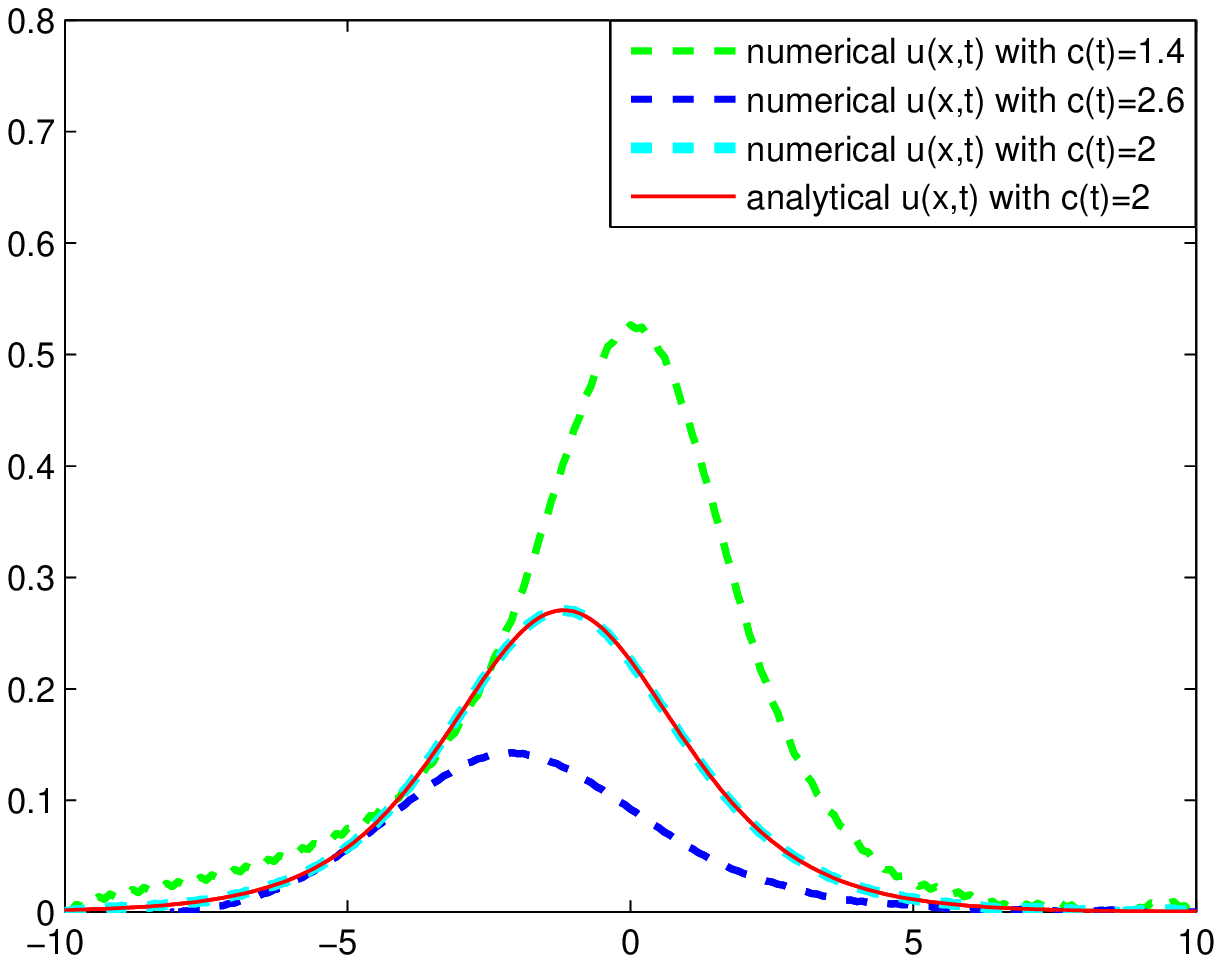}
\vspace{-0.5cm}{\center\hspace{3.7cm}\footnotesize ($a$)
\hspace{8.6cm}($b$)}

\hspace{0cm}\includegraphics[scale=0.5]{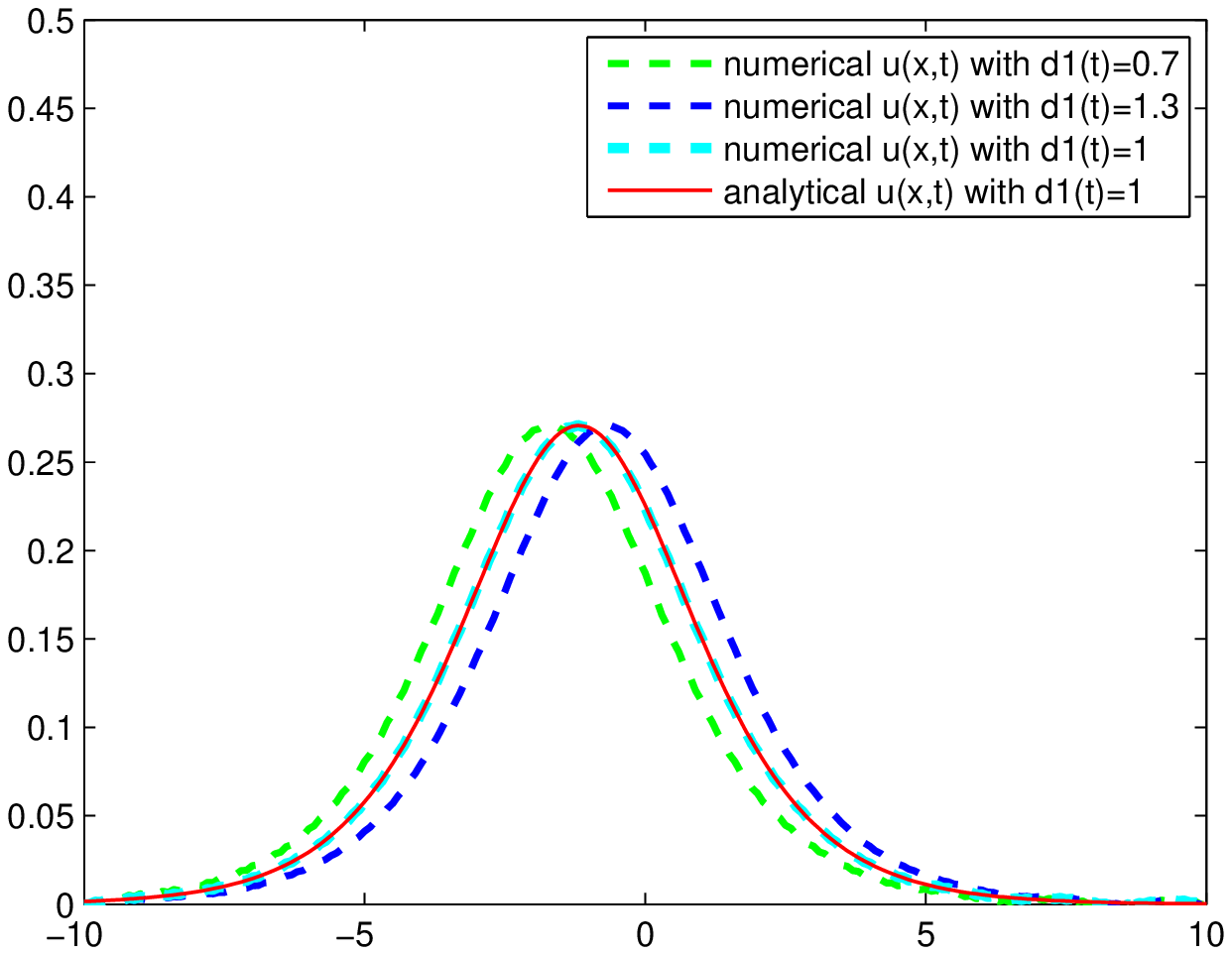}
\hspace{1.5cm}\includegraphics[scale=0.5]{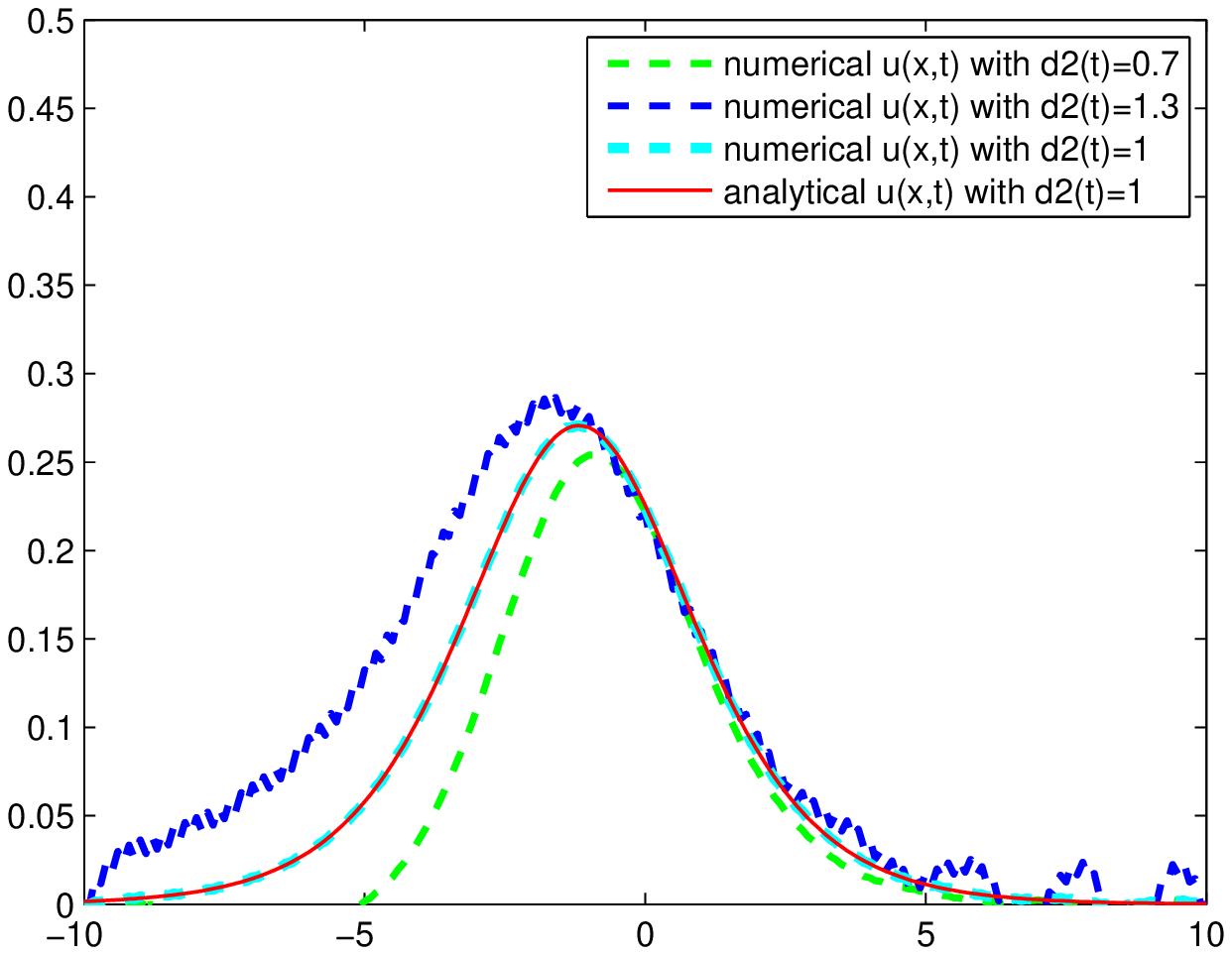}
\vspace{-0.5cm}{\center\hspace{3.7cm}\footnotesize ($c$)
\hspace{8.6cm}($d$)} \figcaption{Numerical simulation with
perturbation on respective coefficients }\label{tu9}
\end{minipage}
\\[\intextsep]

From the view of Fig.9, the consistence between the numerical and
analytical solitonic profiles at $t=1$ is well demonstrated when
constraint~(\ref{cc}) is satisfied.

As shown in Fig.9, the positive (negative) perturbation on
dispersive coefficients $b(t)$ will lead to slight numerical shock
as space expands, with decrease (increase) on velocity and
amplitude. By numerical simulation, coefficient $c(t)$ has an
inhibitory effect on the amplitude of solitons, which is similar to
the analytical discussion under constraint in Section IV. The
decrease on $d_1(t)$ demonstrates slow-down effect on solitons
velocity with the amplitude almost unchanged. The negative
perturbation on $d_2(t)$ results in the decrease of soliton width,
while the positive perturbation can lead to the widening.

\vspace{7mm} \noindent {\Large{\bf IV. Conclusions}}

\vspace{3mm}In this paper, with the nonhomogeneous media taken into
account, we have investigated the nonisospectral and
various-coefficients KdV equation~(\ref{equation}). Under the
general constraint~(\ref{cc}), the DT~(\ref{dtf1}) and~(\ref{dtf2})
is constructed with the adjustments applied to generate the solitons
solutions. By virtue of DT, the multi-soliton solutions are iterated
from the seed solution.

The soliton dynamics might describe different physical phenomena and
benefit the relevant applications in various fields. Therefore, by
virtue of analytical discussion, we discuss the effect of spectral
parameters, coefficients, and initial phases on the solitons
dynamics. The conclusions inferred from the above discussions can be
presented as follow:

(1) As demonstrated in Fig.3, the increase (decrease) of spectral
parameter might induce the growth (fall) of the solitons amplitude
and velocity. With appropriate selection of spectral parameters and
$\chi$, the multi solitons might converge into one at the
intersection.

(2) As shown in Fig.5 and Fig.6, with the proper selection of
line-damping coefficient $c(t)$, we can obtain the space-time
locality of solitons. Besides, the solitons compression (widening)
depends on the sign of coefficient $d_2(t)$.

(3) As illustrated in Fig.7 and Fig.8, by modulating the initial
phases, we can generate the solitons interaction and position
exchange.

Moreover, with the constraint taken into account, the numerical
simulation result in Fig.9 is consistent with the analytical
discussion. With the constraint~(\ref{cc}) perturbed, the effect of
respective coefficient is also studied.

It is demonstrated that the amplitude, width and velocity of
solitons could be modulated by appropriate selection and combination
of the coefficients and parameters. Therefore we can achieve the
purpose of soliton management with explicit conditions taken into
account.

\vspace{7mm} \noindent {\Large{\bf Acknowledgements}}

\vspace{3mm}We express our sincere thanks to Editor, Referees and
all the members of our discussion group for their valuable comments.
This work has been supported by the National Natural Science
Foundation of China under Grant No. 11302014, and by the Fundamental
Research Funds for the Central Universities under Grant Nos.
50100002013105026 and 50100002015105032 (Beijing University of
Aeronautics and Astronautics).

\end{document}